\documentclass[onecolumn,showpacs,preprintnumbers,amsmath,amssymb]{revtex4}
\usepackage{graphicx}
\usepackage{dcolumn}
\usepackage{bm}
\begin{document}
%%%%%%%%%%%%%%%%%%%%%%%%
\newcommand{\hs}{\hspace*{0.5cm}}
\newcommand{\vs}{\vspace*{0.5cm}}
\newcommand{\be}{\begin{equation}}
\newcommand{\ee}{\end{equation}}
\newcommand{\bea}{\begin{eqnarray}}
\newcommand{\eea}{\end{eqnarray}}
\newcommand{\ben}{\begin{enumerate}}
\newcommand{\een}{\end{enumerate}}
\newcommand{\nn}{\nonumber}
\newcommand{\crn}{\nonumber \\}
\newcommand{\non}{\nonumber}
\newcommand{\noi}{\noindent}
\newcommand{\al}{\alpha}
\newcommand{\la}{\lambda}
\newcommand{\bet}{\beta}
\newcommand{\ga}{\gamma}
\newcommand{\va}{\varphi}
\newcommand{\om}{\omega}
\newcommand{\pa}{\partial}
\newcommand{\fr}{\frac}
\newcommand{\bc}{\begin{center}}
\newcommand{\ec}{\end{center}}
\newcommand{\Ga}{\Gamma}
\newcommand{\de}{\delta}
\newcommand{\De}{\Delta}
\newcommand{\ep}{\epsilon}
\newcommand{\varep}{\varepsilon}
\newcommand{\ka}{\kappa}
\newcommand{\La}{\Lambda}
\newcommand{\si}{\sigma}
\newcommand{\Si}{\Sigma}
\newcommand{\ta}{\tau}
\newcommand{\up}{\upsilon}
\newcommand{\Up}{\Upsilon}
\newcommand{\ze}{\zeta}
\newcommand{\ps}{\psi}
\newcommand{\Ps}{\Psi}
\newcommand{\ph}{\phi}
\newcommand{\vph}{\varphi}
\newcommand{\Ph}{\Phi}
\newcommand{\Om}{\Omega}
\def\lappeq{\mathrel{\rlap{\raise.5ex\hbox{$<$}}
{\lower.5ex\hbox{$\sim$}}}}
%%%%%%%%%%%%%%%%%%%%%%%%

\title{$\mbox{SU}(3)_C\otimes \mbox{SU}(3)_L \otimes
\mbox{U}(1)_X$ model with two Higgs triplets}

\author{P. V. Dong}
\email{pvdong@iop.vast.ac.vn}
\author{H. N. Long}
\email{hnlong@iop.vast.ac.vn}
\author{D. T. Nhung}
\affiliation{Institute of Physics, VAST, P. O. Box 429, Bo Ho,
Hanoi 10000, Vietnam}
\author{D. V. Soa}
\email{dvsoa@assoc.iop.vast.ac.vn} \affiliation{Department of
Physics, Hanoi University of Education, Hanoi, Vietnam}

\date{\today}

\begin{abstract}
The $\mathrm {SU}(3)_C\otimes \mathrm {SU}(3)_L \otimes {\mathrm
U}(1)_X$ gauge model with the  minimal scalar sector (two Higgs
triplets) is studied in detail. One of the vacuum  expectation
values $u$  is a source of lepton-number violations and a reason
for the  mixing among the charged gauge bosons - the standard
model $W$ and the bilepton (with $L=2$) gauge bosons as well as
among neutral non-Hermitian  $X^0$ and neutral gauge bosons: the
photon, the $Z$ and the new  $Z'$. Because of these mixings, the
lepton-number violating interactions exist in both charged and
neutral gauge boson sectors. An exact diagonalization of the
neutral gauge boson sector  is derived and bilepton mass splitting
is also given. The lepton-number violation happens only in the
neutrino but not in the charged lepton sector. In this model,
lepton-number changing ($\De L = \pm 2$) processes exist but {\it
only} in the neutrino sector. Constraints on  VEVs of the model
are estimated and $u \simeq \emph{O}(1)\
  \textrm{GeV},
v \simeq v_{weak} =  246 \ \textrm{GeV}$ and $\om  \simeq
\emph{O}(1)\ \textrm{TeV}$.
\end{abstract}


\maketitle

\section{Introduction}
\label{sec:Intro}

The $\mathrm{SU}(3)_C\otimes \mathrm{SU}(2)_L\otimes {\mathrm
U}(1)_Y$ standard model of the strong and electroweak
interactions, with the $\mathrm{SU}(2)_L\otimes {\mathrm U}(1)_Y$
symmetry spontaneously broken down to the ${\mathrm U}(1)_Q$ of
electromagnetism, is an excellent description of the interactions
of elementary particles down to distances in the order of
$10^{-16}\mathrm{cm}$. However it also leaves many striking
features of the physics of our world unexplained. Some of them are
the generation number problem, the electric charge quantization
and the neutrino oscillations \cite{superk} which confirm that
neutrinos are massive and the flavor lepton number is not
conserved. It suggests that it is important to point out the
complete dynamics of Higgs fields. All this requires new
interactions beyond the conventional interactions in the standard
model  (SM).

A very common alternative to solve some of these problems consists
of enlarging the group of gauge symmetry, where the larger group
embeds properly the SM. For instance, the $\mbox{SU}(5)$ grand
unification model \cite{vd} can unify the interactions and
predicts the electric charge quantization, while the group
$\mbox{E}_6$ can also unify the interactions and might explain the
masses of the neutrinos \cite{gr}, and etc. \cite{ps}.
Nevertheless, such models cannot explain the generation number
problem.

A very interesting alternative to explain the origin of
generations comes from the cancelation of chiral anomalies
\cite{agg}. In particular, the models with gauge group $G_{331}=
\mbox{SU}(3)_C\otimes \mbox{SU}(3)_L \otimes \mbox{U}(1)_X$, also
called 3-3-1 models \cite{ppf,flt,long,recent}, arise as a
possible solution to this puzzle, since some of such models
require the three generations in order to cancel chiral anomalies
completely. An additional motivation to study this kind of models
comes from the fact that they can also predict the electric charge
quantization \cite{prs} and the neutrino oscillation
\cite{neu331}.

Such 3-3-1 models have been studied extensively over the last
decade. In one \cite{ppf} of them the three known left-handed
lepton components for each generation are associated to three
$\mathrm{SU}(3)_L$ triplets as $(\nu_l,l,l^c)_L$, where $l^c_L$ is
related to the right-handed isospin singlet of the charged lepton
$l$ in the SM. This model requires that the Higgs sector contains
three scalar triplets and one scalar sextet. In the variant model
\cite{flt,long}  three $\mathrm{SU}(3)_L$ lepton triplets are of
the form $(\nu_l, l, \nu_l^c)_L$, where $\nu_l^c$ is related to
the right-handed component of the neutrino field $\nu_l$ (a model
with right-handed neutrinos). The scalar sector of this model
requires three Higgs triplets, therefore, hereafter we call this
version the 3-3-1 model with three Higgs triplets (331RH3HT). It
is interesting to note that, in the  331RH3HT,  two Higgs triplets
has the same $\mathrm{U}(1)_X$ charge with two neutral components
at their top and bottom. Allowing these neutral components vacuum
expectation values (VEVs), we can reduce number of Higgs triplets
to be two. As a result, the dynamics symmetry breaking also affect
lepton number. Hence it follows that the lepton number is also
broken spontaneously at a high scale of energy. This kind of model
was proposed in Ref.\cite{ponce}, but has not got enough
attention. In Ref.\cite{study}, phenomenology of this model was
presented without mixing between charged gauge bosons as well as
neutral ones.

Phenomenology of the 3-3-1 model  in the version that includes
right-handed neutrinos with two Higgs triplets is a subject of
this study.

The paper is organized as follows. In Sec. II we recall the idea
of constructing the two-Higgs 3-3-1 model.  Secs. III and IV are
devoted to  fermion masses and Higgs potential, respectively. In
Sec V,  masses of charged gauge bosons are given and an exact
diagonalization of neutral ones and their mixings are presented.
Because of the mixings, currents in this model have unusual
features in the neutrino sectors which are presented in Sec. VI.
In Sec. VII, constraints on the parameters of the model and some
phenomena  are sketched. We  outline our main results in the last
section - Sec. VIII.

\section{The particle content}

The particle content in this model which is anomaly free,   is
given as follows: \bea \psi_{iL}&=&\left(%
\begin{array}{c}
  \nu_i \\
  e_i \\
  \nu^c_i \\
\end{array}%
\right)_L\sim \left(3,-\fr 1 3\right), \mbox{ }e_{iR}\sim
(1,-1),\mbox{ }
i=1,2,3,\crn Q_{1L}&=&\left(%
\begin{array}{c}
  u_1 \\
  d_1 \\
  U \\
\end{array}%
\right)_L\sim \left(3,\fr 1 3\right),\mbox{ } Q_{\al L}=\left(%
\begin{array}{c}
  d_\al\\
  -u_\al\\
  D_\al\\
\end{array}%
\right)_L\sim (3^*,0),\mbox{ }\al=2,3,\crn u_{i
R}&\sim&\left(1,\fr 2 3\right), \mbox{ }d_{i R} \sim \left(1,-\fr
1 3\right), \mbox{ }U_{R}\sim \left(1,\fr 2 3\right),\mbox{
}D_{\al R} \sim \left(1,-\fr 1 3\right).\eea Here, the values in
the parentheses denote quantum numbers based on the
$\left(\mbox{SU}(3)_L,\mbox{U}(1)_X\right)$ symmetry. In this
case, the electric charge operator takes a form\be
Q=T_3-\fr{1}{\sqrt{3}}T_8+X,\label{eco}\ee where $T_a$
$(a=1,2,...,8)$ and $X$ stand for $\mbox{SU}(3)_L$ and
$\mbox{U}(1)_X$ charges, respectively.  Electric charges of the
exotic quarks $U$ and $D_\al$  are the same as of the usual
quarks, i.e. $q_{U}=\fr 2 3$ and $q_{D_\al}=-\fr 1 3$.

The $\mathrm{SU}(3)_L\otimes \mathrm{U}(1)_X$ gauge group is
broken spontaneously via two steps. In the first step, it is
embedded in
that of the SM via a Higgs scalar triplet \be \chi=\left(%
\begin{array}{c}
  \chi^0_1 \\
  \chi^-_2 \\
  \chi^0_3 \\
\end{array}%
\right) \sim \left(3,-\fr 1 3\right)\ee acquired with VEV given by
\be
\langle\chi\rangle=\fr{1}{\sqrt{2}}\left(%
\begin{array}{c}
  u \\
  0 \\
  \om \\
\end{array}%
\right).\label{vevc}\ee In the last step, to embed the gauge group
of the SM in $\mathrm{U}(1)_Q$, another Higgs scalar triplet \be
\phi=\left(%
\begin{array}{c}
  \phi^+_1 \\
  \phi^0_2 \\
  \phi^+_3 \\
\end{array}%
\right)\sim \left(3,\fr 2 3\right)\ee is needed with the VEV as
follows
\be \langle\phi\rangle=\fr{1}{\sqrt{2}}\left(%
\begin{array}{c}
  0 \\
  v \\
  0 \\
\end{array}%
\right).\label{vevp}\ee The Yukawa interactions which induce
masses for the fermions can be written in the most general form as
\be
\mathcal{L}_Y=(\mathcal{L}^\chi_Y+\mathcal{L}^\phi_Y)+\mathcal{L}^{\mathrm{mix}}_Y,\ee
where \bea ({\cal L}^{\chi}_Y+{\cal
L}^\phi_Y)&=&h'_{11}\overline{Q}_{1L}\chi
U_{R}+h'_{\al\beta}\overline{Q}_{\al L}\chi^* D_{\beta R}\crn
&&+h^e_{ij}\overline{\psi}_{iL}\phi
e_{jR}+h^\ep_{ij}\ep_{pmn}(\overline{\psi}^c_{iL})_p(\psi_{jL})_m(\phi)_n
+h^d_{1i}\overline{Q}_{1 L}\phi d_{i R}+h^d_{\al
i}\overline{Q}_{\al L}\phi^* u_{iR}+h.c.,\label{y1}\\ {\cal
L}^{\mathrm{mix}}_Y&=&h^u_{1i}\overline{Q}_{1L}\chi
u_{iR}+h^u_{\al i}\overline{Q}_{\al L}\chi^* d_{i
R}+h''_{1\al}\overline{Q}_{1L}\phi D_{\al R}+h''_{\al
1}\overline{Q}_{\al L}\phi^* U_{R}+h.c.\label{y2}\eea The VEV
$\om$ gives mass for the exotic quarks $U$ and $D_\al$, $u$ gives
mass for $u_1, d_{\al}$ quarks, while $v$ gives mass for $u_\al,
d_{1}$ and {\it all} ordinary leptons. In the next section we
provide more details on analysis of fermion masses. As mentioned
above, the VEV $\om $ is responsible for the first step of
symmetry breaking, while the second step is due to $u$ and $v$.
Therefore the  VEVs in this model have to be satisfied the
constraints \be u, v \ll \om . \label{vevcons} \ee

The Yukawa couplings of Eq.(\ref{y1}) possess an extra global
symmetry which implies a new conserved charge $(\mathcal{L})$
through the lepton number $(L)$ by diagonal matrices
\cite{changlong} $L=xT_3+yT_8+\mathcal{L}$. Applying $L$ on a
lepton triplet, the coefficients will be defined \be
L=\fr{4}{\sqrt{3}}T_8+\mathcal{L}.\ee Here, the
$\mathcal{L}$-charges of the fermion and Higgs multiplets can be
obtained by\be \mathcal{L}(\psi_{iL},Q_{1L}, Q_{\al
L},\phi,\chi,e_{iR},u_{iR},d_{iR},U_R,D_{\al R})=\fr 1 3,-\fr 2
3,\fr 2 3,-\fr 2 3,\fr 4 3,1,0,0,-2,2.\ee It is worth  emphasizing
that $\mathcal{L}$ is not broken by VEVs $v,\om$ but by $u$ which
is behind $L(\chi^0_1)=2$ (see also \cite{hml}). This means that
{\it $u$ is a kind of the lepton-number violating parameter}.
Moreover, the Yukawa couplings of (\ref{y2}) also violate
${\mathcal L}$ with $\pm 2$ units which confirm that they are very
small.

\section{\label{app2}Fermion masses}

The fermions gain mass terms via Yukawa interactions given in
(\ref{y1}) and (\ref{y2}).

\subsection{The charged lepton sector}
The charged leptons gain masses via the following Yukawa term\bea
{\cal L}^e_Y=h^e_{ij}\overline{\psi}_{iL} \phi e_{jR}+h.c.\eea
Suppose that the coupling constant $h^e_{ij}$ is diagonal in the
flavor indices $ij$, from which  masses of  the $e, \mu,\tau$ are
followed by \bea m^e=\fr{h^e_{11}v}{\sqrt{2}},\mbox{ }
m^\mu=\fr{h^e_{22}v}{\sqrt{2}},\mbox{
}m^\tau=\fr{h^e_{33}v}{\sqrt{2}}.\eea This is quite similar to
 the SM.

\subsection{The neutrino sector}
The model treats neutrinos with  both left-handed and right-handed
ones. The latter under the $G_{331}$ group are the left-handed
anti-particles transforming as $\mathrm{SU}(3)_L$ triplets in the
same with the former. As a result, this naturally gives rise to an
inverted hierarchy mass pattern and interesting mixing.

At the tree level, the neutrinos gain Dirac masses via \bea{\cal
L}^\ep_Y= h^\ep_{ij}\ep^{p m
n}(\overline{\psi}^c_{iL})_p(\psi_{jL})_m(\phi)_n+h.c.\eea
Hence, the Dirac mass matrix is obtained by\bea M_D=\sqrt{2}v\left(%
\begin{array}{ccc}
  0 & h^\ep_{e\mu} & h^\ep_{e\tau} \\
  -h^\ep_{e\mu} & 0 & h^\ep_{\mu\tau} \\
  -h^\ep_{e\tau} & -h^\ep_{\mu\tau} & 0 \\
\end{array}%
\right),\eea which gives the mass pattern $0,
-m_\nu,m_\nu=iv\sqrt{2(h^{\ep 2}_{e\mu}+h^{\ep 2}_{e\tau}+h^{\ep
2}_{\mu\tau})}$. This is clearly not realistic under the current
data \cite{fl}. However, this pattern is severely changed by the
quantum effect. The most general mass matrix can be written in the
base of $(\nu_e,\nu_\mu,\nu_\tau,\nu^c_e,\nu^c_\mu,\nu^c_\tau)_L$
as
\bea M_{\nu_{LR}}=\left(%
\begin{array}{cc}
  M_{L} & M_{D} \\
  M_D^T & M_{R} \\
\end{array}%
\right),\eea where $M_L$ and $M_R$ arise  from quantum correction.

\subsection{The quark sector}

From the Yukawa terms for the up quarks $(u_1,u_2,u_3,U)$, their
tree level mass matrix is obtained by\be
M_u=-\fr{1}{\sqrt{2}}\left(\begin{array}{cccc} -h^u_{11}u &
-h^u_{12}u & -h^u_{13}u & -h'_{11}u
\\h^{d}_{21}v &
h^{d}_{22}v & h^{d}_{23}v & h''_{21}v
\\ h^{d}_{31}v & h^{d}_{32}v & h^{d}_{33}v & h''_{31}v \\
-h^u_{11}\om & -h^u_{12}\om & -h^u_{13}\om & -h'_{11}\om \\
\end{array}\right).\ee
Here the quarks are mixing due to having the same charges,
however, we can use the way of the SM, i.e. the couplings are here
supposed to be flavor diagonal. Thus, the tree level masses are
derived from \bea m^{c}&=&-h^d_{22}\fr{v}{\sqrt{2}},\mbox{
}m^{t}=-h^d_{33}\fr{v}{\sqrt{2}},\crn
m^{U'}&=&\fr{h'_{11}\om+h^u_{11}u}{\sqrt{2}},\mbox{ }
m^{u'}=0.\eea Note that $u_2$ and $u_3$ are mass eigenstates, thus
they can be  identified  to the  $c$ and $ t$ quarks,
respectively. However, $u$ quark is then mixing with the $U$
exotic quark to give one  massless quark \bea
u'=\fr{h'_{11}u_1-h^u_{11}U}{\sqrt{h^{u2}_{11}+h'^2_{11}}},\eea
and another quark with the mass in the range of $\om$ \bea
U'=\fr{u u_1+\om U}{\sqrt{u^2+\om^2}}.\eea Unlike the usual
331RH3HT, where the third family of quarks should be
discriminating \cite{longvan}, in the  model under consideration
the {\it first} family has to be different from the two others.

Next, the tree level mass matrix of the down quarks
$(d_1,d_2,d_3,D_2,D_3)$  is also followed: \be
M_d=\fr{1}{\sqrt{2}}\left(\begin{array}{ccccc}
h^d_{11}v & h^d_{12}v  & h^d_{13}v  & h''_{12}v  & h''_{13}v \\
h^u_{21}u & h^u_{22}u  & h^u_{23}u  & h'_{22}u  & h'_{23}u \\
h^u_{31}u & h^u_{32}u  & h^u_{33}u  & h'_{32}u  & h'_{33}u \\
h^u_{21}\om & h^u_{22}\om  & h^u_{23}\om  & h'_{22}\om  & h'_{23}\om \\
h^u_{31}\om & h^u_{32}\om  & h^u_{33}\om  & h'_{32}\om  & h'_{33}\om \\
\end{array}\right). \ee To get masses, it can be supposed that the
couplings are flavor diagonal. Thus, it is easy to obtain \bea
m^d&=&h^d_{11}\fr{v}{\sqrt{2}},\mbox{ }m^{s'}=0,\mbox{
}m^{b'}=0,\crn m^{S'}&=&\fr{h^u_{22}u+h'_{22}\om}{\sqrt{2}},\mbox{
}m^{B'}=\fr{h^u_{33}u+h'_{33}\om}{\sqrt{2}}.\label{23}\eea Here
$d_1$ is a mass eigenstate which is identified with $d$ quark. The
quarks $d_2$ and $D_2$ ($d_3$ and $D_3$) are mixing to give one
massless quark
 \bea
s'=\fr{h'_{22}d_2-h^u_{22}D_2}{\sqrt{h^{u2}_{22}+h'^2_{22}}}\mbox
{
}\left(b'=\fr{h'_{33}d_3-h^u_{33}D_3}{\sqrt{h^{u2}_{33}+h'^2_{33}}}\right),\eea
and another quark with the mass in the order of $\om$ (see
Eq.(\ref{23}))\bea S'=\fr{u d_2+\om D_2}{\sqrt{u^2+\om^2}} \mbox{
} \left(B'=\fr{u d_3+\om D_3}{\sqrt{u^2+\om^2}}\right).\eea The
two remaining states with masses of the range $\om$ belong to
exotic quarks. The masslessness of the $u_1,d_2,d_3$ quarks calls
radiative corrections and the interested reader can  see also  in
Ref.\cite{study}.

\section{Higgs potential}

In this model, the most general Higgs potential has very simple
form \bea V(\chi,\phi) &=& \mu_1^2 \chi^\dag \chi + \mu_2^2
\phi^\dag \phi + \la_1 ( \chi^\dag \chi)^2 + \la_2 ( \phi^\dag
\phi)^2\crn &  & + \la_3 ( \chi^\dag \chi)( \phi^\dag \phi) +
\la_4 ( \chi^\dag \phi)( \phi^\dag \chi). \label{poten} \eea Note
that there is no trilinear scalar coupling and this makes the
Higgs potential much simpler than the previous ones
\cite{long98,changlong} and closer to that of the SM. The analysis
in Ref.\cite{ponce} shows that after symmetry breaking, there are
eight Goldstone bosons and four physical scalar fields. One of two
physical neutral scalars is the SM  Higgs boson.

To break spontaneously the symmetry, the Higgs vacuums are not
$\mbox{SU}(3)_L\otimes \mbox{U}(1)_X$ singlets. Thus, non-zero
values of $\chi$ and $\phi$ at the minimum value of $V(\chi,\phi)$
can be easily obtained by\bea \chi^+\chi&=&\fr{\lambda_3\mu^2_2
-2\lambda_2\mu^2_1}{4\lambda_1\lambda_2-\lambda^2_3}
\equiv\fr{u^2+\om^2}{2},\label{vev1}\\
\phi^+\phi&=&\fr{\lambda_3\mu^2_1
-2\lambda_1\mu^2_2}{4\lambda_1\lambda_2-\lambda^2_3}
\equiv\fr{v^2}{2}.\label{vev2}\eea It is worth  noting that any
other choice of $u,\om$ for the vacuum value of $\chi$ satisfying
(\ref{vev1}) gives the same physics because it is related to
(\ref{vevc}) by an $\mbox{SU}(3)_L\otimes \mbox{U}(1)_X$
transformation. Thus, in general case we assume that $u\neq 0$.

\section{Gauge bosons}
The covariant derivative of a triplet is given by \bea D_\mu &=&
\pa_\mu-igT_aW_{a\mu}-ig_X T_9 X B_\mu \crn
&\equiv&\pa_\mu-i\mathcal{P}_\mu, \eea where the gauge fields
$W_a$ and $B$ transform as the adjoint representations of
$\mathrm{SU}(3)_L$ and $\mathrm{U}(1)_X$, respectively, and the
corresponding gauge coupling constants $g$, $g_X$. Moreover,
$T_9=\fr{1}{\sqrt{6}}\mathrm{diag}(1,1,1)$ is fixed so that the
relation $\mbox{Tr}(T_{a'} T_{b'})=\fr{1}{2}\delta_{a'b'}$
$(a',b'=1,2,...,9)$ is satisfied. The $\mathcal{P}_\mu$ matrix
appeared in the above covariant derivative is rewritten in a
convenient form \bea
\fr{g}{2}\left(%
\begin{array}{ccc}
  W_{3\mu}+\fr{1}{\sqrt{3}}W_{8\mu}+t\sqrt{\fr 2 3}XB_\mu
  & \sqrt{2} W'^+_\mu & \sqrt{2}X'^0_\mu \\
  \sqrt{2}W'^-_\mu & -W_{3\mu}+\fr{1}{\sqrt{3}}W_{8\mu}+
  t\sqrt{\fr 2 3}X B_\mu & \sqrt{2}Y'^-_\mu \\
  \sqrt{2}X'^{0*}_\mu & \sqrt{2}Y'^+_\mu &
  -\fr{2}{\sqrt{3}}W_{8\mu}+t\sqrt{\fr 2 3}X B_\mu \\
\end{array}%
\right),\eea where $t\equiv g_X/g$. Let us denote the following
combinations \bea W'^{\pm} _\mu &\equiv& \fr{W_{1\mu}\mp
iW_{2\mu}}{\sqrt{2}},\crn Y'^\mp_\mu &\equiv& \fr{W_{6\mu}\mp
iW_{7\mu}}{\sqrt{2}}, \crn X'^0_\mu &\equiv&
\fr{W_{4\mu}-iW_{5\mu}}{\sqrt{2}} \eea having defined charges
under the generators of the $\mathrm{SU}(3)_L$ group. For the sake
of convenience in  further reading, we note that, $W_4$ and $W_5$
are pure real and imaginary parts of $ X'^0_\mu$ and $
X'^{0*}_\mu$, respectively \bea W_{4\mu} & = & \fr{1}{\sqrt{2}} (
X'^0_\mu + X'^{0*}_\mu),\crn W_{5\mu}& = & \fr{i}{\sqrt{2}} (
X'^0_\mu - X'^{0*}_\mu).\label{w4xx}\eea

The masses of the gauge bosons in this model are followed from
\bea {\cal
L}^{\mathrm{GB}}_{\mathrm{mass}}&=&(D_\mu\langle\phi\rangle)^+
(D^\mu\langle\phi\rangle)+ (D_\mu\langle\chi\rangle)^+
(D^\mu\langle\chi\rangle)\crn &=&\fr{g^2}{4}(u^2+v^2)W'^-_\mu
W'^{+\mu}+\fr{g^2}{4}(\om^2+v^2)Y'^-_\mu
Y'^{+\mu}+\fr{g^2u\om}{4}(W'^-_\mu Y'^{+\mu}+Y'^-_\mu
W'^{+\mu})\crn & & +
\fr{g^2v^2}{8}\left(-W_{3\mu}+\fr{1}{\sqrt{3}}W_{8\mu}+t\fr 2 3
\sqrt{\fr 2 3} B_\mu\right)^2\crn & & +
\fr{g^2u^2}{8}\left(W_{3\mu}+\fr{1}{\sqrt{3}}W_{8\mu}-t\fr 1 3
\sqrt{\fr 2 3}B_\mu
\right)^2+\fr{g^2\om^2}{8}\left(-\fr{2}{\sqrt{3}}W_{8\mu}-t\fr 1
3\sqrt{\fr 2 3} B_\mu\right)^2\crn & & + \fr{g^2 u
\om}{4\sqrt{2}}\left(W_{3\mu}+\fr{1}{\sqrt{3}}W_{8\mu}-t \fr 1 3
\sqrt{\fr 2 3}B_\mu\right)\left(X'^{0\mu}+X'^{0*\mu}\right)\crn &
& + \fr{g^2 u \om}{4\sqrt{2}}\left(-\fr{2}{\sqrt{3}}W_{8\mu}-t\fr
1 3 \sqrt{\fr 2 3}
B_\mu\right)\left(X'^{0\mu}+X'^{0*\mu}\right)\crn &  & +
\fr{g^2}{16}(u^2+\om^2)\left\{(X'^0_\mu+X'^{0*}_\mu)^2+
[i(X'^0_\mu-X'^{0*}_\mu)]^2\right\}.\label{ngbmassm}\eea

The combinations  $W'$ and $Y'$ are mixing via\bea {\cal
L}^{\mathrm{CG}}_{\mathrm{mass}}&=&\fr{g^2}{4}(W'^-_\mu,Y'^-_\mu)\left(%
\begin{array}{cc}
  u^2+v^2 & u\om \\
  u\om & \om^2+v^2 \\
\end{array}%
\right)\left(%
\begin{array}{c}
  W'^{+\mu} \\
  Y'^{+\mu} \\
\end{array}%
\right).\eea Diagonalizing this mass matrix, we get {\it physical}
charged gauge bosons
 \bea W^-_\mu &=& \cos\theta
W'^-_\mu+\sin\theta Y'^-_\mu ,\crn Y^-_\mu &=& -\sin\theta
W'^-_\mu+\cos\theta Y'^-_\mu,\eea where the mixing angle is
defined by \be \tan\theta=\fr{u}{\om}.\ee The
mass eigenvalues are  \bea M^2_{W}&=&\fr{g^2v^2}{4},\label{massw}\\
M^2_{Y}&=&\fr{g^2}{4}(u^2+v^2+\om^2).\label{massy}\eea Because of
the constraints in (\ref{vevcons}), the following remarks are in
order: \ben\item $\theta$ should be very small, and then $W_\mu
\simeq W'_\mu, Y_\mu \simeq Y'_\mu$. \item  $v \simeq v_{weak} =
246$ GeV due to identification of $W$ as the $W$ boson in the
SM.\een

Next, from (\ref{ngbmassm}), the $W_5$ gains mass as follows \be
M^2_{W_5}=\fr{g^2}{4}(\om^2+u^2).\ee

Finally, there is a mixing among $W_3, W_8, B, W_4$ components. In
the basis of these elements, the mass matrix is given by
\bea M^2=\fr{g^2}{4}\left(%
\begin{array}{cccc}
  u^2+v^2 & \fr{u^2-v^2}{\sqrt{3}} & -\fr{2t}{3\sqrt{6}}(u^2+2v^2) & 2u\om \\
  \fr{u^2-v^2}{\sqrt{3}} & \fr{1}{3}(4\om^2+u^2+v^2) &
  \fr{\sqrt{2}t}{9}(2\om^2-u^2+2v^2)
   & -\fr{2}{\sqrt{3}}u\om \\
  -\fr{2t}{3\sqrt{6}}(u^2+2v^2) & \fr{\sqrt{2}t}{9}(2\om^2-u^2+2v^2)
  & \fr{2t^2}{27}(\om^2+u^2+4v^2)
  & -\fr{8t}{3\sqrt{6}}u\om \\
  2u\om & -\fr{2}{\sqrt{3}}u\om & -\fr{8t}{3\sqrt{6}}u\om & u^2+\om^2 \\
\end{array}%
\right). \label{nmass}\eea Note that the mass Lagrangian in this
case has the form \bea {\cal
L}^{\mathrm{NG}}_{\mathrm{mass}}&=&\fr 1 2 V^T M^2 V,\crn V^T
&\equiv& (W_3, W_8, B, W_4).\eea In the limit $u\rightarrow 0$,
$W_{4}$ does not mix with $W_{3\mu}, W_{8\mu}, B_\mu$. In the
general case $u \neq 0$,  the  mass matrix in (\ref{nmass})
contains  two {\it exact eigenvalues} such as  \bea M^2_\ga &=&0,
\crn M^2_{W'_4}&=&\fr{g^2}{4}(\om^2+u^2).\label{massx}\eea Thus
the $W'_4$ and $W_5$ components have the same mass, and this
conclusion {\it  contradicts the previous analysis in}
Ref.\cite{ponce}. With this result, we should  identify the
combination of $W'_4$ and $W_5$ \be \sqrt{2} X_\mu^0 = W'_{4\mu} -
i W_{5\mu} \label{chbln} \ee as {\it physical} neutral {\it
non-Hermitian}  gauge boson. The subscript $0$ denotes neutrality
of gauge boson $X$. However, in the following, this subscript may
be dropped. This boson caries lepton number two, hence it is the
bilepton like those in the usual 331RH3HT. From (\ref{massw}),
(\ref{massy}) and (\ref{massx}), it follows an interesting
relation between  the bilepton masses similar to the law of
Pythagoras \bea M^2_{Y}&=& M^2_{X}+M^2_{W}. \label{massrel} \eea
Thus the charged bilepton $Y$ is slightly heavier than the neutral
one $X$. Remind that the similar relation in the 331RH3HT is
\cite{il}:  $ | M_Y^2 - M_X^2 | \leq m_W^2$.

Now we turn to the eigenstate question. The  eigenstates
corresponding to the two values in (\ref{massx}) are determined as
follows
\bea A_\mu=\fr{1}{\sqrt{18+4t^2}}\left(%
\begin{array}{c}
  \sqrt{3}t \\
  -t \\
  3\sqrt{2} \\
  0 \\
\end{array}%
\right), \mbox{    } W'_{4\mu}=\fr{1}{\sqrt{1+4\tan^22\theta}}\left(%
\begin{array}{c}
  \tan2\theta \\
  \sqrt{3}\tan2\theta \\
  0 \\
  1 \\
\end{array}%
\right).\eea To embed this model in the effective theory at the
low energy we follow an appropriate  method without Higgs in Ref.
\cite{ld,mohapatra}, where the photon field couples with the
lepton by strength \be {\cal
L}^{\mathrm{EM}}_{\mathrm{int}}=-\fr{\sqrt{3}g_X}{\sqrt{18+4t^2}}\overline{l}\ga^\mu
l A_\mu.\ee Therefore the coefficient of the electromagnetic
coupling constant can be identified as \be
\fr{\sqrt{3}g_X}{\sqrt{18+4t^2}} =e\ee Using continuation of the
gauge coupling constant $g$ of $\mathrm{SU}(3)_L$ at the
spontaneous symmetry breaking point \be
g=g[\mathrm{SU}(2)_L]=\fr{e}{s_W}\ee from which it follows \be
t=\fr{3\sqrt{2}s_W}{\sqrt{3-4s^2_W}}.\ee The eigenstates are now
rewritten as follows\bea A_\mu &=& s_W
W_{3\mu}+c_W\left(-\fr{t_W}{\sqrt{3}}
W_{8\mu}+\sqrt{1-\fr{t^2_W}{3}}B_\mu\right),\crn
W'_{4\mu}&=&\fr{t_{2\theta}}{\sqrt{1+4t^2_{2\theta}}}W_{3\mu}+
\fr{\sqrt{3}t_{2\theta}}{\sqrt{1+4t^2_{2\theta}}}W_{8\mu}
+\fr{1}{\sqrt{1+4t^2_{2\theta}}}W_{4\mu},\eea where we have
denoted $s_W\equiv\sin\theta_W$, $t_{2\theta}\equiv\tan2\theta$,
and so forth.

The diagonalization of the mass matrix is done via three steps. In
the first step, in  the base of $(A_\mu,Z_\mu,Z'_\mu,W_{4\mu})$,
the two remaining gauge vectors are given by \bea Z_\mu&=& c_W
W_{3\mu}-s_W\left(-\fr{t_W}{\sqrt{3}}
W_{8\mu}+\sqrt{1-\fr{t^2_W}{3}}B_\mu\right),\crn Z'_\mu &=&
\sqrt{1-\fr{t^2_W}{3}} W_{8\mu}+\fr{t_W}{\sqrt{3}}B_\mu.\eea In
this basis, the mass matrix $M^2$ becomes \bea M'^2=\fr{g^2}{4}\left(%
\begin{array}{cccc}
  0 & 0 & 0 & 0 \\
  0 & \fr{u^2 + v^2}{c^2_W} & \fr{c_{2W}u^2-v^2}{c^2_W
  \sqrt{3-4s^2_W}} & \fr{2 u \om}{c_W} \\
  0 & \fr{c_{2W}u^2-v^2}{c^2_W\sqrt{3-4s^2_W}} &
  \fr{v^2+4c^4_W\om^2+c^2_{2W}u^2}{c^2_W(3-4s^2_W)} &
  -\fr{2 u \om}{c_W\sqrt{3-4s^2_W}} \\
  0 & \fr{2 u \om}{c_W} & -\fr{2 u \om}{c_W\sqrt{3-4s^2_W}} & u^2+\om^2 \\
\end{array}%
\right).\eea Also, in the limit $u\rightarrow 0$, $W_{4\mu}$ does
not mix with $Z_{\mu},Z'_{\mu}$. The eigenstate $W'_{4\mu}$ is now
defined by \bea
W'_{4\mu}=\fr{t_{2\theta}}{c_W\sqrt{1+4t^2_{2\theta}}}Z_\mu+
\fr{\sqrt{4c^2_W-1}t_{2\theta}}{c_W\sqrt{1+
4t^2_{2\theta}}}Z'_\mu+\fr{1}{\sqrt{1+4t^2_{2\theta}}}W_{4\mu}.\eea
We turn to the second step. To see explicitly that the following
basis is orthogonal and normalized, let us put\be
s_{\theta'}\equiv\fr{t_{2\theta}}{c_W\sqrt{1+4t^2_{2\theta}}},\label{htmix}\ee
which leads to \be W'_{4\mu}=s_{\theta'} Z_\mu+
c_{\theta'}\left[t_{\theta'}\sqrt{4c^2_W-1}Z'_\mu+
\sqrt{1-t^2_{\theta'}(4c^2_W-1)}W_{4\mu}\right].\label{thetapr}\ee
Note that the mixing angle in this step $\theta'$ is the same
order as  the mixing angle in the charged gauge boson sector.
Taking into account \cite{pdg} $s^2_W \simeq 0.231$, from
(\ref{htmix})   we get $s_{\theta'}\simeq 2.28 s_{\theta}$. It is
now easy to choose two remaining gauge vectors \bea
\mathcal{Z}_{\mu}&=&c_{\theta'} Z_\mu-
s_{\theta'}\left[t_{\theta'}\sqrt{4c^2_W-1}Z'_\mu+
\sqrt{1-t^2_{\theta'}(4c^2_W-1)}W_{4\mu}\right],\crn
\mathcal{Z}'_{\mu}&=&\sqrt{1-t^2_{\theta'}(4c^2_W-1)}Z'_\mu-
t_{\theta'}\sqrt{4c^2_W-1}W_{4\mu}.\eea Therefore, in the base of
$(A_\mu,\mathcal{Z}_{\mu},{\cal Z}'_{\mu}$,$W'_{4\mu})$  the mass
matrix $M'^2$  has a quasi-diagonal form  \bea M''^2=\left(%
\begin{array}{cccc}
  0 & 0 & 0 & 0 \\
  0 & m^2_{\mathcal{Z}} & m^2_{\mathcal{Z}\mathcal{Z}'} & 0 \\
  0 & m^2_{\mathcal{Z}\mathcal{Z}'} & m^2_{\mathcal{Z}'} & 0 \\
  0 & 0 & 0 & \fr{g^2}{4}(u^2+\om^2) \\
\end{array}%
\right)\label{fmassmatrix}\eea with \bea
m^2_{\mathcal{Z}}&=&\fr{g^2[(1+3t^2_{2\theta})u^2+(1+4t^2_{2\theta})v^2-
t^2_{2\theta}\om^2]}{4[c^2_W+(3-4s^2_W)t^2_{2\theta}]},\crn
m^2_{\mathcal{Z}\mathcal{Z}'}&=&\fr{g^2\sqrt{1+4t^2_{2\theta}}
\left\{[c_{2W}+(3-4s^2_W)t^2_{2\theta}]u^2-v^2-(3-4s^2_W)t^2_{2\theta}
\om^2\right\}}{4\sqrt{3-4s^2_W}[c^2_W+(3-4s^2_W)t^2_{2\theta}]},\crn
m^2_{\mathcal{Z}'}&=&\fr{g^2\left\{[c^2_{2W}+(3-4s^2_{2W})t^2_{2\theta}
]u^2+v^2+[4c^4_W+(1+4c^2_W)(3-4s^2_W)t^2_{2\theta}]\om^2\right\}}
{4(3-4s^2_W)[c^2_W+(3-4s^2)t^2_{2\theta}]}.\eea

In the last step, it is trivial  to diagonalize the  mass matrix
in (\ref{fmassmatrix}). The two remaining mass eigenstates are
given by \bea Z^1_\mu &=& c_\va {\cal Z}_{\mu}-s_\va
\mathcal{Z}'_{\mu},\crn Z^2_\mu &=& s_\va {\cal Z}_{\mu}+c_\va
\mathcal{Z}'_{\mu},\eea where the mixing angle $\va$ between
$\mathcal{Z}$ and $\mathcal{Z}'$ is defined by \be
t_{2\va}=\fr{\sqrt{(3-4s^2_W)(1+4t^2_{2\theta})}\left\{[c_{2W}+(3-4s^2_W)t^2_{2\theta}]u^2-
v^2-(3-4s^2_W)t^2_{2\theta}\om^2\right\}}{[2s^4_W-1+(8s^4_W-2s^2_W-3)
t^2_{2\theta}]u^2-
[c_{2W}+2(3-4s^2_W)t^2_{2\theta}]v^2+[2c^4_W+(8s^4_W+9c_{2W})t^2_{2\theta}]\om^2}.\ee
The physical mass eigenvalues are defined  by\bea
M^2_{Z^1}&=&\fr{c^2_W(u^2+\om^2)+v^2-\sqrt{[c^2_W
(u^2+\om^2)+v^2]^2+(3-4s^2_W)(3u^2\om^2-
u^2v^2-v^2\om^2)}}{2g^{-2}(3-4s^2_W)},\crn
M^2_{Z^2}&=&\fr{c^2_W(u^2+\om^2)+v^2+\sqrt{[c^2_W
(u^2+\om^2)+v^2]^2+(3-4s^2_W)(3u^2\om^2
-u^2v^2-v^2\om^2)}}{2g^{-2}(3-4s^2_W)}.\nn\eea Because of the
condition (\ref{vevcons}), the angle $\va$ has to be  very
small\be
t_{2\va}\simeq-\fr{\sqrt{3-4s^2_W}[v^2+(11-14s^2_W)u^2]}{2c^4_W\om^2}.\ee
In this approximation,  the above physical states have masses \bea
M^2_{Z^1}&\simeq& \fr{g^2}{4c^2_W}(v^2-3u^2),\label{massz1}\\
M^2_{Z^2}&\simeq& \fr{g^2c^2_W\om^2}{3-4s^2_W}.\eea Consequently,
$Z^1$ can be identified as the $Z$ boson in the SM, and $Z^2$
being the new neutral (Hermitian) gauge boson. It is important to
note that in the limit  $u\rightarrow 0$ the mixing angle $\va$
between $\mathcal{Z}$ and $\mathcal{Z}'$ is always non-vanishing.
This differs from the mixing angle $\theta$ between the $W$ boson
of the SM and the singly-charged bilepton $Y$. Phenomenology of
the mentioned mixing is quite similar to the $W_L - W_R$ mixing in
the left-right symmetric model based on the $\textrm{SU}(2)_R
\otimes \textrm{SU}(2)_L \otimes \textrm{U}(1)_{B-L}$ group (the
interested reader can find in \cite{mohapatra}).

\section{Currents}

The interaction among fermions with gauge bosons arises from part
\bea i \bar{\psi}\ga_\mu D^\mu \psi =  \textrm{kinematic terms} +
H^{\mathrm{CC}} + H^{\mathrm{NC}}. \label{lagcurrent} \eea

\subsection{Charged currents}
Despite neutrality, the gauge bosons $X^0$, $X^{0*}$ belong to
this section by their nature.  Because of the mixing among the SM
$W$ boson and the charged bilepton $Y$ as well as among ($X^0 +
X^{0*}$) with $(W_3, W_8, B)$,   the new terms exist as follows
\bea H^{\mathrm{CC}}=\fr{g}{\sqrt{2}}\left(J^{\mu-}_W W^+_\mu +
J^{\mu-}_Y Y^+_\mu + J^{\mu 0*}_X X^{0}_\mu + h.c.\right) \eea
where \bea J^{\mu-}_W&=&c_\theta (\overline{\nu}_{iL}\ga^\mu
e_{iL}+\overline{u}_{iL}\ga^\mu d_{iL})+s_\theta
(\overline{\nu}^c_{iL}\ga^\mu
e_{iL}+\overline{U}_{L}\ga^\mu d_{1L}+\overline{u}_{\al L}\ga^\mu D_{\al L}),\label{dongw}\\
J^{\mu-}_Y&=&c_\theta (\overline{\nu}^c_{iL}\ga^\mu
e_{iL}+\overline{U}_{L}\ga^\mu d_{1L}+\overline{u}_{\al L}\ga^\mu
D_{\al L})-s_\theta (\overline{\nu}_{iL} \ga^\mu
e_{iL}+\overline{u}_{i L}\ga^\mu d_{i L}),\label{dongy}\\
J^{\mu 0*}_X &=& (1-t^2_{2\theta})(\overline{\nu}_{iL}\ga^\mu
\nu^c_{i L}+\overline{u}_{1L}\ga^\mu U_{L}-\overline{D}_{\al
L}\ga^\mu d_{\al
L})-t^2_{2\theta}(\overline{\nu}^c_{iL}\ga^\mu\nu_{iL}+\overline{U}_L\ga^\mu
u_{1L}-\overline{d}_{\al L}\ga^\mu D_{\al L})\crn
&&+\fr{t_{2\theta}}{\sqrt{1+4t^2_{2\theta}}}(\overline{\nu}_i\ga^\mu
\nu_i+\overline{u}_{1L}\ga^\mu u_{1L}-\overline{U}_L\ga^\mu
U_L-\overline{d}_{\al L}\ga^\mu d_{\al L}+\overline{D}_{\al
L}\ga^\mu D_{\al L}).\label{dongx} \eea

Comparing with the charged currents in the usual
331RH3HT\cite{long}  we get the following discrepances \ben
\item The second term in (\ref{dongw})
\item The second term in (\ref{dongy})
\item The second and the third terms in (\ref{dongx})
\een All mentioned above interactions are  lepton-number violating
and weak (proportional to $\sin \theta$ or its square  $\sin^2
\theta$). However, these couplings lead to lepton-number
violations only in the neutrino sector.

\subsection{Neutral currents}

As before, in this model, a real part of the non-Hermitian neutral
$X'^0$ mixes with the real neutral ones such as $Z$ and $Z'$. This
gives the {\it unusual} term as follows
 \bea H^{\mathrm{NC}} = e
A^\mu J^{\mathrm{EM}}_\mu + {\mathcal L}^{\mathrm{NC} } +
{\mathcal L}^{\mathrm{NC} }_{\mathrm{unnormal}}. \label{ncurrent}
\eea

Despite the mixing among $W_3, W_8, B, W_4$, the electromagnetic
interactions {\it remain the same as} in the SM and the usual
331RH3HT, i.e. \be J^{\mathrm{EM}}_\mu = \sum_f q_f \bar{f}
\ga_\mu f,
 \label{ecurrent} \ee
where $f$ runs among all the fermions of the model.

Interactions of the neutral currents with  fermions have a common
form \bea {\mathcal
L}^{\mathrm{NC}}&=&\fr{g}{2c_W}\overline{f}\ga^\mu \left[g_{k
V}(f)-g_{k A}(f)\ga^5\right] f Z^k_\mu,\mbox{  } k= 1,
2\label{normal}\eea where \bea
g_{1V}(f)&=&\fr{c_\va\left\{T_3(f_L)-3t^2_{2\theta}
X(f_L)+[(3-8s^2_W)t^2_{2\theta}-2s^2_W]Q(f)\right\}
}{\sqrt{(1+4t^2_{2\theta})[1+(3-t^2_W)t^2_{2\theta}]}} \crn
&&-\fr{s_\va [(4c^2_W-1)T_3(f_L)+3c^2_W X(f_L)-(3-5s^2_W)Q(f)]}
{\sqrt{(4c^2_W-1)[1+(3-t^2_W)t^2_{2\theta}]}},\\
g_{1A}(f)&=&\fr{c_\va
[T_3(f_L)-3t^2_{2\theta}(X-Q)(f_L)]}{\sqrt{(1+4t^2_{2\theta})
[1+(3-t^2_W)t^2_{2\theta}]}}-\fr{s_\va
[(4c^2_W-1)T_3(f_L)+3c^2_W(X-Q)(f_L)]}{
\sqrt{(4c^2_W-1)[1+(3-t^2_W)t^2_{2\theta}]}},\\
g_{2V}(f)&=&g_{1V}(f)(c_\va\rightarrow s_\va, s_\va\rightarrow
-c_\va),\mbox{  } g_{2A}(f)=g_{1A}(f)(c_\va\rightarrow s_\va,
s_\va\rightarrow -c_\va).\eea

Here $T_3(f)$ and $Q(f)$ are, respectively, the third component of
the weak isospin and the charge of the fermion $f$. Note that
isospin for the $SU(2)$ fermion singlet (in the bottom of
triplets) vanishes: $T_3(f)= 0$. The values of  $g_{1V}(f)$,
$g_{1A}(f) $ and $g_{2V}(f)$, $g_{2A}(f)$ are listed in Table
\ref{tab1} and Table \ref{tab2}.

\begin{table}
\caption{\label{tab1} The $Z^1_\mu \rightarrow \bar{f} f$
couplings.}
\begin{ruledtabular}
\begin{tabular}{lcc}
$f$ \hs & \hs  $g_{1V}(f)$ \hs & \hs  $g_{1A}(f)$ \\ \hline \\
$\nu_e,\nu_\mu,\nu_\tau$\hs & \hs $\fr{c_\va -s_\va
\sqrt{(4c^2_W-1)(1+4t^2_{2\theta})}}
{2\sqrt{(1+4t^2_{2\theta})[1+(3-t^2_W)t^2_{2\theta}]}}$&
$\fr{c_\va \sqrt{(4c^2_W-1)(1+4t^2_{2\theta})}+s_\va}
{2\sqrt{(4c^2_W-1)[1+(3-t^2_W)t^2_{2\theta}]}}$\\ \\
$e,\mu,\tau$ \hs & \hs
$\fr{(3-4c^2_W)[c_\va\sqrt{(4c^2_W-1)(1+4t^2_{2\theta})}+s_\va]}
{2\sqrt{(4c^2_W-1)[1+(3-t^2_W)t^2_{2\theta}]}}$\hs  & \hs
$-\fr{c_\va\sqrt{(4c^2_W-1)(1+4t^2_{2\theta})}+s_\va}
{2\sqrt{(4c^2_W-1)[1+(3-t^2_W)t^2_{2\theta}]}}$\\ \\
$u$\hs & \hs$\fr{c_\va
\sqrt{4c^2_W-1}[3(1+2t^2_{2\theta})-8s^2_W(1+4t^2_{2\theta})]
-s_\va(3+2s^2_W)\sqrt{1+4t^2_{2\theta}}}{6\sqrt{(4c^2_W-1)(1+4t^2_{2\theta})
[1+(3-t^2_W)t^2_{2\theta}]}}$ &
$\fr{c_\va\sqrt{4c^2_W-1}(1+2t^2_{2\theta})-s_\va
c_{2W}\sqrt{1+4t^2_{2\theta}}}{2\sqrt{(4c^2_W-1)(1+4t^2_{2\theta})
[1+(3-t^2_W)t^2_{2\theta}]}}$\\ \\
$d$ \hs &\hs
$\fr{(1-4c^2_W)[c_\va\sqrt{(4c^2_W-1)(1+4t^2_{2\theta})}+s_\va]}
{6\sqrt{(4c^2_W-1)[1+(3-t^2_W)t^2_{2\theta}]}}$ &
$-\fr{c_\va\sqrt{(4c^2_W-1)(1+4t^2_{2\theta})}+s_\va}
{2\sqrt{(4c^2_W-1)[1+(3-t^2_W)t^2_{2\theta}]}}$\\ \\
$c,t$\hs &\hs $\fr{(3-8s^2_W)[c_\va\sqrt{(4c^2_W-1)
(1+4t^2_{2\theta})}+s_\va]}{6\sqrt{(4c^2_W-1)
[1+(3-t^2_W)t^2_{2\theta}]}}$ &
$\fr{c_\va\sqrt{(4c^2_W-1)(1+4t^2_{2\theta})}+s_\va}
{2\sqrt{(4c^2_W-1)[1+(3-t^2_W)t^2_{2\theta}]}}$\\ \\
$s,b$\hs &\hs $\fr{c_\va
\sqrt{4c^2_W-1}[(1-4c^2_W)(1+4t^2_{2\theta})+6t^2_{2\theta}]+s_\va(1+2c^2_W)
\sqrt{1+4t^2_{2\theta}}}{6\sqrt{(4c^2_W-1)
(1+4t^2_{2\theta})[1+(3-t^2_W)t^2_{2\theta}]}}$ & $-\fr{c_\va
\sqrt{4c^2_W-1}(1+2t^2_{2\theta})-s_\va
c_{2W}\sqrt{1+4t^2_{2\theta}}}
{2\sqrt{(4c^2_W-1)(1+4t^2_{2\theta})[1+(3-t^2_W)t^2_{2\theta}]}}$\\
\\
$U$ \hs &\hs $\fr{c_\va
\sqrt{4c^2_W-1}[3t^2_{2\theta}-4s^2_W(1+4t^2_{2\theta})] +s_\va
(3-7s^2_W)\sqrt{1+4t^2_{2\theta}}}{3\sqrt{(4c^2_W-1)(1+4t^2_{2\theta})
[1+(3-t^2_W)t^2_{2\theta}]}}$ & $\fr{c_\va
\sqrt{4c^2_W-1}t^2_{2\theta}+s_\va
c^2_W\sqrt{1+4t^2_{2\theta}}}{\sqrt{(4c^2_W-1)(1+4t^2_{2\theta})[1+(3-t^2_W)t^2_{2\theta}]}}$\\
\\
$D_2,D_3$ \hs &\hs $\fr{c_\va
\sqrt{4c^2_W-1}[2s^2_W(1+4t^2_{2\theta})-3t^2_{2\theta}]
-s_\va(3-5s^2_W)\sqrt{1+4t^2_{2\theta}}}{3\sqrt{(4c^2_W-1)(1+4t^2_{2\theta})
[1+(3-t^2_W)t^2_{2\theta}]}}$ & $-\fr{c_\va
\sqrt{4c^2_W-1}t^2_{2\theta}+s_\va
c^2_W\sqrt{1+4t^2_{2\theta}}}{\sqrt{(4c^2_W-1)(1+4t^2_{2\theta})
[1+(3-t^2_W)t^2_{2\theta}]}}$
\end{tabular}
\end{ruledtabular}
\end{table}

\begin{table}
\caption{\label{tab2} The $Z^2_\mu \rightarrow \bar{f} f$
couplings.}
\begin{ruledtabular}
\begin{tabular}{lcc}
$f$ \hs & \hs  $g_{2V}(f)$ \hs & \hs  $g_{2A}(f)$ \\ \hline \\
$\nu_e,\nu_\mu,\nu_\tau$\hs &
$\fr{s_\va+c_\va\sqrt{(4c^2_W-1)(1+4t^2_{2\theta})}}
{2\sqrt{(1+4t^2_{2\theta})[1+(3-t^2_W)t^2_{2\theta}]}}$&
$\fr{s_\va\sqrt{(4c^2_W-1)(1+4t^2_{2\theta})}-c_\va}
{2\sqrt{(4c^2_W-1)[1+(3-t^2_W)t^2_{2\theta}]}}$\\ \\
$e,\mu,\tau$ & \hs
$\fr{(3-4c^2_W)[s_\va\sqrt{(4c^2_W-1)(1+4t^2_{2\theta})}-c_\va]}
{2\sqrt{(4c^2_W-1)[1+(3-t^2_W)t^2_{2\theta}]}}$\hs  & \hs
$-\fr{s_\va\sqrt{(4c^2_W-1)(1+4t^2_{2\theta})}-c_\va}
{2\sqrt{(4c^2_W-1)[1+(3-t^2_W)t^2_{2\theta}]}}$\\
\\$u$
&\hs$\fr{s_\va
\sqrt{4c^2_W-1}[3(1+2t^2_{2\theta})-8s^2_W(1+4t^2_{2\theta})]
+c_\va(3+2s^2_W)\sqrt{1+4t^2_{2\theta}}}{6\sqrt{(4c^2_W-1)(1+4t^2_{2\theta})
[1+(3-t^2_W)t^2_{2\theta}]}}$ &
$\fr{s_\va\sqrt{4c^2_W-1}(1+2t^2_{2\theta})+c_\va
c_{2W}\sqrt{1+4t^2_{2\theta}}}{2\sqrt{(4c^2_W-1)(1+4t^2_{2\theta})
[1+(3-t^2_W)t^2_{2\theta}]}}$\\ \\
$d$ &
$\fr{(1-4c^2_W)[s_\va\sqrt{(4c^2_W-1)(1+4t^2_{2\theta})}-c_\va]}
{6\sqrt{(4c^2_W-1)[1+(3-t^2_W)t^2_{2\theta}]}}$ &
$-\fr{s_\va\sqrt{(4c^2_W-1)(1+4t^2_{2\theta})}-c_\va}
{2\sqrt{(4c^2_W-1)[1+(3-t^2_W)t^2_{2\theta}]}}$\\ \\
$c,t$ & $\fr{(3-8s^2_W)[s_\va\sqrt{(4c^2_W-1)
(1+4t^2_{2\theta})}-c_\va]}{6\sqrt{(4c^2_W-1)
[1+(3-t^2_W)t^2_{2\theta}]}}$ &
$\fr{s_\va\sqrt{(4c^2_W-1)(1+4t^2_{2\theta})}-c_\va}
{2\sqrt{(4c^2_W-1)[1+(3-t^2_W)t^2_{2\theta}]}}$\\ \\
$s,b$ & $\fr{s_\va
\sqrt{4c^2_W-1}[(1-4c^2_W)(1+4t^2_{2\theta})+6t^2_{2\theta}]-
c_\va(1+2c^2_W)\sqrt{1+4t^2_{2\theta}}}{6\sqrt{(4c^2_W-1)
(1+4t^2_{2\theta})[1+(3-t^2_W)t^2_{2\theta}]}}$ & $-\fr{s_\va
\sqrt{4c^2_W-1}(1+2t^2_{2\theta})+c_\va
c_{2W}\sqrt{1+4t^2_{2\theta}}}
{2\sqrt{(4c^2_W-1)(1+4t^2_{2\theta})[1+(3-t^2_W)t^2_{2\theta}]}}$\\
\\
$U$ & $\fr{s_\va
\sqrt{4c^2_W-1}[3t^2_{2\theta}-4s^2_W(1+4t^2_{2\theta})]
-c_\va(3-7s^2_W)\sqrt{1+4t^2_{2\theta}}}{3\sqrt{(4c^2_W-1)(1+4t^2_{2\theta})
[1+(3-t^2_W)t^2_{2\theta}]}}$ & $\fr{s_\va
\sqrt{4c^2_W-1}t^2_{2\theta}-c_\va
c^2_W\sqrt{1+4t^2_{2\theta}}}{\sqrt{(4c^2_W-1)(1+4t^2_{2\theta})[1+(3-t^2_W)t^2_{2\theta}]}}$\\
\\
$D_2,D_3$ & $\fr{s_\va
\sqrt{4c^2_W-1}[2s^2_W(1+4t^2_{2\theta})-3t^2_{2\theta}]
+c_\va(3-5s^2_W)\sqrt{1+4t^2_{2\theta}}}{3\sqrt{(4c^2_W-1)(1+4t^2_{2\theta})
[1+(3-t^2_W)t^2_{2\theta}]}}$ & $-\fr{s_\va
\sqrt{4c^2_W-1}t^2_{2\theta}-c_\va
c^2_W\sqrt{1+4t^2_{2\theta}}}{\sqrt{(4c^2_W-1)(1+4t^2_{2\theta})
[1+(3-t^2_W)t^2_{2\theta}]}}$
\end{tabular}
\end{ruledtabular}
\end{table}

Because of the above-mentioned mixing, the lepton-number violating
interactions mediated by neutral gauge bosons $Z^1$ and $Z^2$
exist in the {\it neutrino  and the exotic quark sectors}
 \bea {\mathcal L}^{\mathrm{NC} }_{\mathrm{unnormal}}&=&-\fr{g
 t_{2\theta}g_{kV}(\nu)}{2}(\overline{\nu}_{iL}\ga^\mu \nu^c_{i
L}+\overline{u}_{1L}\ga^\mu U_{L}-\overline{D}_{\al L}\ga^\mu
d_{\al L}) Z^k_\mu + h.c. \label{un} \eea Again, these
interactions are very weak and proportional to  $\sin \theta$.
From (\ref{dongw}) - (\ref{dongx}) and (\ref{un}) we conclude that
all lepton-number violating interactions are expressed in the
terms dependent only  in  the mixing angle between the charged
gauge bosons.

\section{Phenomenology}

\begin{figure*}[htbp]
\begin{center}
\includegraphics[width=10cm,height=7cm]{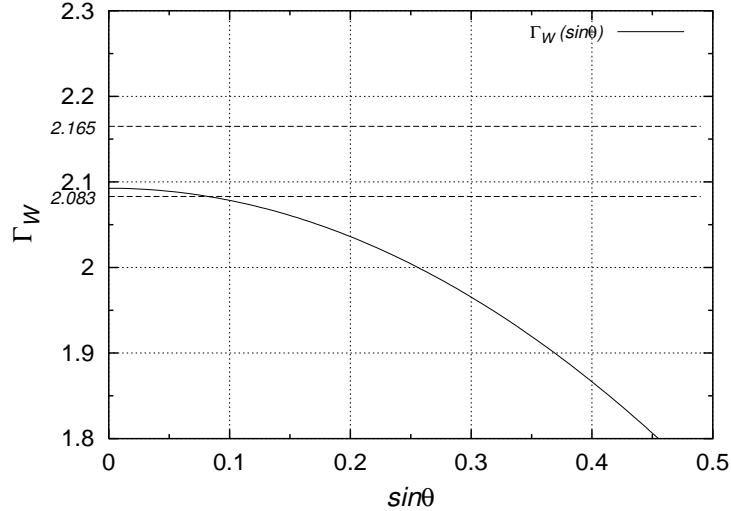}
\caption{\label{wsin}{$W$ width as function of $\sin \theta$, and
the horizontal lines are an upper and a lower limit. }}
\end{center}
\end{figure*}

First of all we should  find some constraints on the parameters of
the model. There are many ways to get constraints on the mixing
angle $\theta$ and the charged bilepton mass $M_Y$. Below we
present a simple one. In our model, the $W$ boson has the
following {\it normal main} decay modes: \bea W^-  & \rightarrow &
l\  \tilde{\nu}_l\  (l = e,\mu,\tau),\crn
 &  \searrow & u^c
d, u^c s, u^c b,  (u\rightarrow c),\label{decaywn}\eea
 which are the same as
in the SM and in the 331RH3HT. Beside the above  modes, there are
additional ones which are lepton-number violating $(\De L = 2)$ -
the model's  specific feature   \be W^- \rightarrow l \ \nu_l \ (l
= e,\mu,\tau). \label{decaywan}\ee It is easy to compute the tree
level decay widths as follows \cite{bardin} \bea
\Ga^{\mathrm{Born}}(W\rightarrow l\ \tilde{\nu}_l)&=&\fr{g^2
c^2_{\theta}}{8}\fr{M_W}{6\pi}(1-x)(1-\fr{x}{2}-\fr{x^2}{2})\simeq
\fr{c^2_\theta \al M_W}{12 s^2_W},\crn
\Ga^{\mathrm{Born}}(W\rightarrow l\ \nu_l)&=&\fr{g^2
s^2_{\theta}}{8}\fr{M_W}{6\pi}(1-x)(1-\fr{x}{2}-\fr{x^2}{2})\simeq
\fr{s^2_\theta \al M_W}{12 s^2_W},\hs x\equiv m^2_l/M^2_W,\crn
\sum_{\mathrm{color}}\Ga^{\mathrm{Born}}(W\rightarrow u^c_i
d_j)&=&\fr{3g^2
c^2_{\theta}}{8}\fr{M_W}{6\pi}|V_{ij}|^2\left[1-2(x+\overline{x})+(x-\overline{x})^2\right]^{\fr
{1}{2}}\crn
&&\times\left[1-\fr{x+\overline{x}}{2}-\fr{(x-\overline{x})^2}{2}\right]\simeq
\fr{c^2_\theta \al M_W}{4 s^2_W}|V_{ij}|^2,\hs x \equiv
m^2_{d_j}/M^2_W,\mbox{ } \overline{x} \equiv
m^2_{u^c_i}/M^2_W.\label{qcd}\eea QCD radiative corrections modify
Eq.(\ref{qcd}) by a multiplicative factor \cite{pdg,bardin}\be
\de_{\mathrm{QCQ}}=
1+\al_s(M_Z)/\pi+1.409\al^2_s/\pi^2-12.77\al^3_s/\pi^3\simeq
1.04,\ee which is estimated from $\al_s(M_Z)\simeq 0.12138$. All
the state masses can be ignored, the predicted total width for $W$
decay into fermions is \bea \Ga^{\mathrm{tot}}_W=1.04\fr{\al
M_W}{2s^2_W}(1-s^2_{\theta})+\fr{\al M_W}{4s^2_W}.\eea Taking
$\al(M_Z) \simeq 1/128$, $M_W=80.425\mathrm{GeV}$, $s^2_W=0.2312$
and $\Ga^{\mathrm{tot}}_W=2.124\pm 0.041 \mathrm{GeV}$ \cite{pdg},
in Fig.\ref{wsin}, we have plotted   $ \Ga^{\mathrm{tot}}_W$ as
function of $s_{\theta}$.
From the figure we get an upper  limit: $\sin \theta \leq 0.08$.

\begin{figure}
\includegraphics{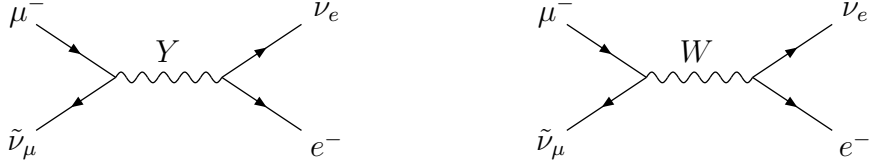}
\caption{\label{fig:fig1} Feynman diagram for the wrong muon decay
$\mu^-\rightarrow e^- \nu_e \tilde{\nu}_\mu$.}
\end{figure}

Since one of the VEVs is closely to the those in the SM:  $v
\simeq v_{weak}= 246 $ GeV, therefore only two free VEVs exist  in
the considering model, namely $u$ and $\om$. The bilepton mass
limit can be obtained from the ``wrong"  muon decay
 \be \mu^- \rightarrow e^- \nu_e \tilde{\nu}_\mu
\label{muondecayw} \ee  mediated,  at the tree level, by both the
SM $W$ and the singly-charged  bilepton $Y_\mu$ (see
Fig.\ref{fig:fig1}). Remind that in the 331RH3HT, at the lowest
order, this decay is mediated only by the singly-charged bilepton
$Y$. In our case,  the second diagram in Fig.\ref{fig:fig1} gives
main contribution.
Taking into account of the famous experimental data \cite{pdg}
 \be R_{muon} \equiv
\fr{\Ga(\mu^- \rightarrow e^- \nu_e \tilde{\nu}_\mu)}{\Ga(\mu^-
\rightarrow e^- \tilde{\nu}_e \nu_\mu)}
 < 1.2 \% \hs  \textrm{90 \% \ CL} \label{wrdecayrat}
 \ee
 we get the constraint: $ R_{muon} \simeq \fr{M_W^4}{M_Y^4}$.
 Therefore, it follows that $M_Y \geq 230 $ GeV.

However, the stronger  bilepton mass bound has been derived from
consideration of experimental limit on  lepton-number violating
charged lepton decays \cite{tullyjoshi} of 440 GeV.

In the case of $u \rightarrow 0$, analyzing the $Z$ decay width
\cite{study}, the $Z - Z'$ mixing angle is constrained by $-0.0015
\leq  \varphi \leq 0.001$. From atomic parity violation in cesium,
bounds for mass of the new exotic $Z'$ and the $Z - Z'$ mixing
angles, again in the limit $u\rightarrow 0$, are given
\cite{study} \be -0.00156 \leq \varphi \leq 0.00105, \hs M_{Z_2}
\geq 2.1 \ \textrm{TeV} \label{boundz}\ee These values coincide
with the bounds in the usual 331RH3HT \cite{longtrung}.

For our purpose we consider the $\rho$ parameter - one of the most
important quantities of the SM, having a leading contribution in
terms of the $T$ parameter, is very useful to get the new-physics
effects. It is well-known relation between $\rho$ and $T$
parameter \be \rho = 1 + \al T \label{rhot} \ee In the usual
3331RH3HT, $T$ gets contribution from the oblique correction and
the $Z-Z'$ mixing \cite{il} \bea T_{RHN} &= &T_{Z Z'} +
T_{oblique},\eea where $ T_{Z Z'} \simeq \fr{\tan^2
\varphi}{\al}\left( \fr{M^2_{Z_2}}{M^2_{Z_1}} -1 \right)$ is
negligible for $M_{Z'}$ less than 1 TeV, $T_{oblique}$ depends on
masses of  the top quark and the SM  Higgs boson. Again at  the
tree level and the limit (\ref{vevcons}), from (\ref{massw}) and
(\ref{massz1}) we get an expression for the $\rho$ parameter in
the considering model
 \be \rho = \fr{M^2_{W}}{c^2_W
M^2_{Z^1}}=\fr{v^2}{v^2-3u^2} \simeq 1 + \fr{3u^2}{v^2}.
\label{rhopar} \ee Note that formula (\ref{rhopar}) has only one
free parameter $u$, since   $v$ is very close to the VEV in the
SM. Neglecting the contribution from the usual 331RH3HT and taking
into account the experimental data \cite{pdg} $\rho = 0.9987 \pm
0.0016 $ we get the constraint on $u$ parameter by $ \fr u v \leq
0.01$ which leads to $ u \leq 2.46 $ GeV. This means  that $u$ is
much smaller than $v$, as expected.

It seems that the $\rho$ parameter, at the tree level, in this
model, is favorable to be bigger unit and this is similar to the
case of the models contained  heavy $Z'$ \cite{luo}.

\begin{figure}
\includegraphics{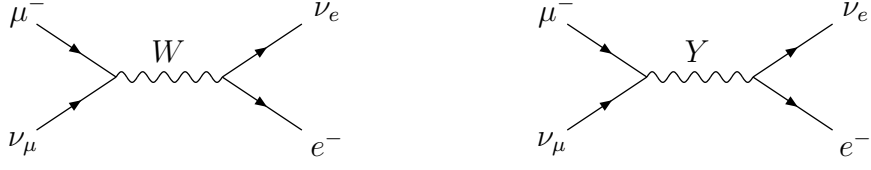}
\caption{\label{fig:fig2} Feynman diagram for $\mu^-\rightarrow
e^- \nu_e \widetilde{\nu}_\mu$.}
\end{figure}

\begin{figure}
\includegraphics{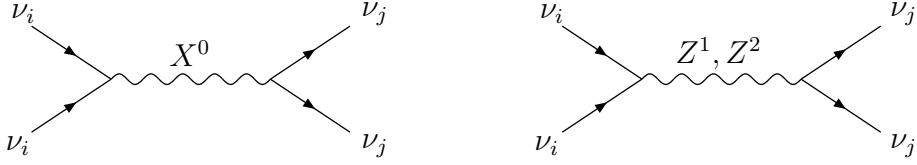}
\caption{\label{fig:fig3} Feynman diagram for
$\nu_i\nu_i\rightarrow \nu_j\nu_j \mbox{ }(i\neq j =e,\mu,\tau)$.}
\end{figure}

The interesting new physics compared with other 3-3-1 models is
the neutrino physics. Due to lepton-number violating couplings we
have the following interesting consequences: \ben
\item {\it Processes with $\De L = \pm 2$}\\
From the charged currents we have the following lepton-number
violating $\De L = \pm 2$ decays such as \bea \mu^- \rightarrow
e^- \nu_e \nu_\mu, \hs \mu^- \rightarrow e^- \tilde{\nu}_e
\tilde{\nu}_\mu, \hs ( \mu \ \textrm{can be replaced by  } \ \tau)
\label{muondecay1} \eea in which both the SM $W$ boson and charged
bilepton $Y_\mu^-$ are in intermediate states (see Fig.
\ref{fig:fig2}).
Here the main contribution arises from the first diagram.
 Note that the wrong muon decay violates only {\it family}
 lepton-number, i.e. $\De L = 0$, but not lepton-number at all as in (\ref{muondecay1}).
 The decay rates are given by \be R_{rare} \equiv
\fr{\Ga(\mu^- \rightarrow e^- \nu_e \nu_\mu)}{\Ga(\mu^-
\rightarrow e^- \tilde{\nu}_e \nu_\mu)} = \fr{\Ga(\mu^-
\rightarrow e^- \tilde{\nu}_e \tilde{\nu}_\mu )}{\Ga(\mu^-
\rightarrow e^- \tilde{\nu}_e \nu_\mu)} \simeq s_\theta^2
\label{decayrat}
 \ee
Taking $s_\theta = 0.08$, we get $R_{rare} \simeq 6 \times
10^{-3}$. This rate is the same as the  wrong  muon decay one.
Interesting to note that, the family lepton-number violating
processes \be \nu_i \nu_i \rightarrow \nu_j \nu_j, \ (i \neq
j)\label{netross}\ee
 are mediated not only by the non-Hermitian
bilepton $X$ but also by the Hermitian neutral $Z^1, Z^2$ (see
Fig.\ref{fig:fig3}).

The first diagram in Fig.\ref{fig:fig3} exists also in the
331RH3HT, but the second one does not appear there.

\item {\it  Lepton-number violating kaon decays}\\
 Next, let us consider the  lepton-number violating   decay \cite{pdg}
\be K^+ \rightarrow \pi^0 + e^+ \tilde{\nu}_e\  < 3 \times 10^{-3}
\ \textrm{at} \  90 \% \ \textrm{CL}
 \ee
This decay can be explained in the considering model  as the
subprocess given below \be \tilde{s} \rightarrow \tilde{u} + e^+
\tilde{\nu}_e. \ee
 This process is mediated by the SM $W$ boson
and the charged bilepton $Y$. Amplitude of the considered process
is proportional to $\sin_\theta$ \be M (\tilde{s} \rightarrow
\tilde{u} + e^+ \tilde{\nu}_e) \simeq \fr{\sin 2 \theta}{ 2 M_W^2}
\left( 1 - \fr{m_W^2}{M_Y^2}\right) \label{radecay}\ee Next, let
us consider the ``normal decay" \cite{pdg} \be K^+ \rightarrow
\pi^0 + e^+ \nu_e\  \   (4.87 \pm 0.06)\ \% \ee with amplitude \be
M (\tilde{s} \rightarrow \tilde{u} + e^+ \nu_e) \simeq \fr{1}{
M_W^2} \label{radecayn}\ee From (\ref{radecay}) and
(\ref{radecayn}) we get \be R_{kaon} \equiv \fr{\Ga (\tilde{s}
\rightarrow \tilde{u} + e^+ \tilde{\nu}_e)}{\Ga (\tilde{s}
\rightarrow \tilde{u} + e^+ \nu_e)} \simeq \sin^2 \theta . \ee

In the framework of this model, we derive  the following decay
modes with rates \bea  R_{kaon} =
 \fr{\Ga (K^+\rightarrow \pi^0 + e^+
\tilde{\nu}_e)}{\Ga (K^+ \rightarrow \pi^0 + e^+ \nu_e)} \simeq
\fr{\Ga (K^+ \rightarrow \pi^0 + \mu^+ \tilde{\nu}_\mu)}{\Ga (K^+
\rightarrow \pi^0 + \mu^+ \nu_\mu)} \simeq \sin^2 \theta \leq 6
\times 10^{-3}. \eea Note that the similar lepton-number violating
processes exist in the
 $\textrm{SU}(2)_R
\otimes \textrm{SU}(2)_L \otimes \textrm{U}(1)_{B-L}$
 model (for details,
see Ref.\cite{mohapatra}).

\item {\it Neutrino Majorana masses}\\
At the one-loop level neutrinos get Majorana masses via diagram in
which all the charged $W, Y$ and the neutral $Z$ and $Z'$ gauge
bosons give contributions. However, the contributions from the
above-mentioned fields give the terms diagonal in flavor basis.
Fortunately, the Higgs scalar with lepton number two gives
interesting term leading to the neutrino oscillation. This result
will be presented elsewhere.

\een

\section{Conclusions}
\label{sec:conclusions}

In this paper we have presented the 3-3-1 model with the minimal
scalar sector (only two Higgs triplets). This version belongs to
the 3-3-1 model without exotic charges (charges of the exotic
quarks are $\fr 2 3$ and $-\fr 1 3$).  The spontaneous symmetry
breakdown is achieved with only two Higgs triplets.  One of the
vacuum  expectation values $u$  is a source of lepton-number
violations and a reason for the  mixing between the charged gauge
bosons - the standard model $W$ and the singly-charged bilepton
gauge bosons as well as between neutral non-Hermitian $X^0$ and
neutral gauge bosons: the photon, the $Z$ and the new  exotic
$Z'$. At the tree level, masses of the charged gauge bosons
satisfy the law of Pythagoras $ M_Y^2 = M_X^2 + M_W^2$ and in the
limit $\om \gg u, v$, the $\rho$ parameter gets additional
contribution dependent only on $\fr u v$. Thus, this leads to $u
\ll v$, and there are three quite different scales for the VEVs of
the model: one is very small $u \simeq \emph{O}(1)  $ GeV - a
lepton-number violating parameter, the second $v$ is close to the
SM one : $v\simeq v_{weak} = 246$ GeV and the last is in the range
of new physics scale about $\emph{O}(1)$ TeV.

In difference with  the usual 331RH3HT, in this model the first
family of quarks should be distinctive of the two others.

The exact diagonalization of the neutral gauge boson sector  is
derived. Because of the parameter $u$, the lepton-number violation
 happens only in neutrino
but not in charged lepton sector. It is interesting to note that
despite the mentioned above mixing, the electromagnetic current
remains unchanged.  In this model, the lepton-number changing
($\De L = \pm 2$) processes exist but only in the neutrino sector.
Consequently, neutrinos get Majorana masses at the one-loop level.

It is worth  mentioning on  the advantage of the considered model:
the new mixing angle between the charged gauge bosons $\theta$ is
connected with one of the VEVs $u$ - the parameter of
lepton-number violations. There is no new  parameter, but it
contains very simple Higgs sector, hence the significant number of
free parameters is reduced.

The model contains new kinds of interactions in the neutrino
sector. Hence neutrino physics in this model is very rich. It
deserves further studies.

\section*{Acknowledgments}

This work was supported in part by National Council for
Natural Sciences of Vietnam.\\[0.3cm]

\appendix

\section{Mixing matrix of neutral gauge bosons}
For the sake of convenience in practical calculations, we need the
mixing matrix
\be \left(%
\begin{array}{c}
  W_3 \\
  W_8 \\
  B \\
  W_4 \\
\end{array}%
\right) =\mathrm{U}\left(%
\begin{array}{c}
  A \\
  Z^1 \\
  Z^2 \\
  W'_{4} \\
\end{array}%
\right),
 \ee where
 \be U= \left(%
\begin{array}{cccc}
  s_W & c_\va c_{\theta'}c_W  & s_\va c_{\theta'}c_W  & s_{\theta'}c_W \\
  -\fr{s_W}{\sqrt{3}} & \fr{c_\va(s^2_W-3c^2_Ws^2_{\theta'})
  -s_\va\sqrt{(1-4s^2_{\theta'}c^2_W)(4c^2_W-1)}}{\sqrt{3}c_Wc_{\theta'}} &
  \fr{s_\va(s^2_W-3c^2_Ws^2_{\theta'})+
  c_\va\sqrt{(1-4s^2_{\theta'}c^2_W)(4c^2_W-1)}}{\sqrt{3}c_Wc_{\theta'}} & \sqrt{3}s_{\theta'}c_W \\
  \fr{\sqrt{4c^2_W-1}}{\sqrt{3}} & -\fr{t_W(c_\va\sqrt{4c^2_W-1}
  +s_\va\sqrt{1-4s^2_{\theta'}c^2_W})}{\sqrt{3}c_{\theta'}} & -\fr{t_W(s_\va\sqrt{4c^2_W-1}
 -c_\va\sqrt{1-4s^2_{\theta'}c^2_W})}{\sqrt{3}c_{\theta'}}  & 0 \\
  0 & -t_{\theta'}(c_\va\sqrt{1-4s^2_{\theta'}c^2_W}
  -s_\va\sqrt{4c^2_W-1}) & -t_{\theta'}(s_\va\sqrt{1-4s^2_{\theta'}c^2_W}
  +c_\va\sqrt{4c^2_W-1}) & \sqrt{1-4s^2_{\theta'}c^2_W} \\
\end{array}%
\right)\nn.\ee Here we have denoted  $s_{\theta'}=
\fr{t_{2\theta}}{c_W\sqrt{1+4t^2_{2\theta}}}$.

%%%%%%%%%%%%%%%%%%%%%%%%%%%%%%%%%%%%%%%%%%%%%%%%%%%%%%

\end{document}